\documentclass{ai4x}

\IACpaperyear{2025} 
\IAClocation{Singapore} 
\IACdate{8--11 July 2025} 

\title{Sparse Autoencoders Reveal Interpretable Structure in Small Gene Language Models}

\hypersetup{
    pdfkeywords={AI, gene language models, sparse autoencoders}
}

\IACauthor{Haoxiang Guan}{0009-0001-4933-9748}{1}{0}
\IACauthor{Jiyan He}{0009-0003-4539-1826}{1}{0}
\IACauthor{Jie Zhang}{0000-0002-4230-1077}{2}{2}

\IACauthoraffiliation{University of Science and Technology of China, Hefei, Anhui, China}
\IACauthoraffiliation{CFAR and IHPC,  Agency for Science, Technology and Research, Singapore, \normalfont{\authormail{zhang\_jie@cfar.a-star.edu.sg}}}

\makeatletter
\def\@maketitle{%
  \newpage
  \begin{center}%
    {\large \textbf{{\@title}} \par}%
    \vskip 1.5em%
    {\large
      \lineskip .5em%
      \newcounter{authnum}%
      \setcounter{authnum}{0}
      \whileboolexpr
      { test {\ifnumcomp{\value{authnum}}{<}{\theauthcount}} }%
        {\stepcounter{authnum}%
            \csdef{this-author}{\normalsize\textbf{\csuse{iac@author\theauthnum}}}%
            {\ifnumcomp{\csuse{iac@ispresentingauthor\theauthnum}}{=}{1}{\underline{\csuse{this-author}}}{\csuse{this-author}}}\textsuperscript{%
                \ifcsempty{iac@orcid\theauthnum}{}%
                          {\orcidlink{\csuse{iac@orcid\theauthnum}}}%
            \@alph{\csuse{iac@affiliationidx\theauthnum}}}%
      	{\ifnumcomp{\value{authnum}}{=}{\theauthcount}{}{,\ }}%
	    }%
    }%
  \end{center}%
  \newcounter{affnum}
    \setcounter{affnum}{0}
    \whileboolexpr
      { test {\ifnumcomp{\value{affnum}}{<}{\theaffiliationcount}} }%
      {\stepcounter{affnum}%
		\textsuperscript{\@alph{\theaffnum}} \normalsize\textit{\csuse{iac@affiliation\theaffnum}}\par%
		\vskip 1.5ex%
	  }
    \setcounter{authnum}{0}
    \whileboolexpr
      { test {\ifnumcomp{\value{authnum}}{<}{\theauthcount}} }%
      {\stepcounter{authnum}%
		\ifnumcomp{\csuse{iac@ispresentingauthor\theauthnum}}{=}{1}{* Presenting author}{}\par%
		 \setcounter{authnum}{99}
	  } 
      \vskip 2.5ex
  }%
\makeatother

\begin{document}
\maketitle
\thispagestyle{fancy} 

\section{Introduction}
\label{sec:introduction}
Sparse autoencoders (SAEs) have recently emerged as a powerful tool for interpreting the internal representations of large language models (LLMs), revealing latent latent features with semantical meaning~\cite{cunningham2023sparseautoencodershighlyinterpretable}. This interpretability has also proven valuable in biological domains:
applying SAEs to protein language models uncovered meaningful features related to protein structure and function~\cite{Simon2024.11.14.623630}. More recently, SAEs have been used to analyze genomics-focused models such as Evo 2~\cite{Brixi2025.02.18.638918}, identifying interpretable features in gene sequences. However, it remains \textbf{\textit{unclear}} whether SAEs can extract meaningful representations from small gene language models, which have fewer parameters and potentially less expressive capacity. To address it, we propose applying SAEs to the activations of a small gene language model.
We demonstrate that even small-scale models encode biologically relevant genomic features, such as transcription factor binding motifs, that SAEs can effectively uncover.  Our findings suggest that compact gene language models are capable of learning structured genomic representations, and that SAEs offer a scalable approach for interpreting gene models across various model sizes.

\section{Methods}
\label{sec:substantial}
\begin{figure*}[htp]
\centering
\includegraphics[width=\textwidth]{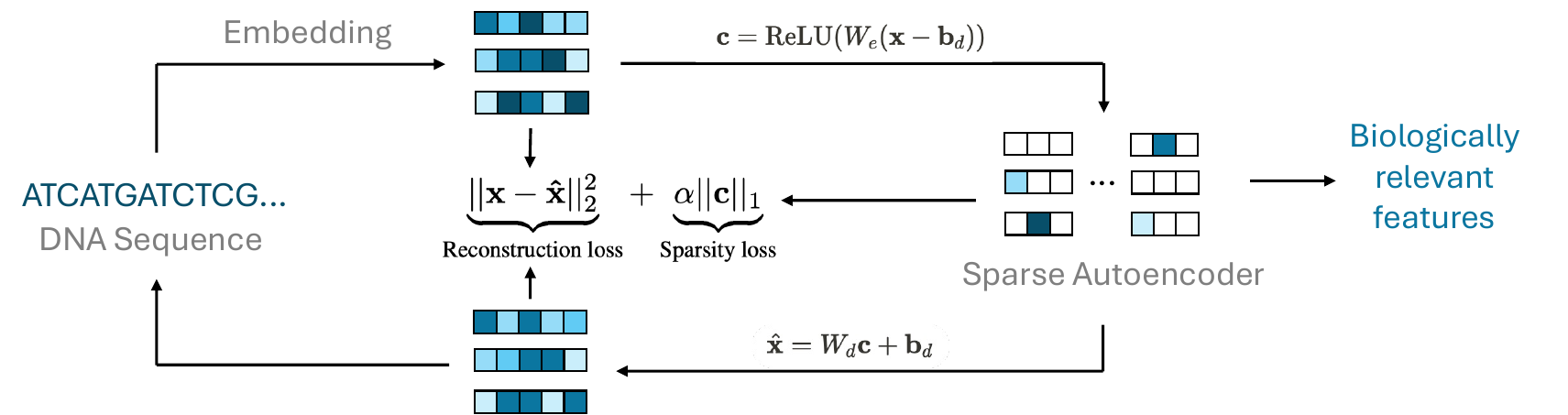}
\caption{Overall pipeline for training SAEs on genomes, followed by identifying biologically relevant features.}
\label{fig:pipeline}
\end{figure*}

To uncover interpretable structures in small gene language models, we trained sparse autoencoders (SAEs) on embeddings derived from HyenaDNA-small-32k~\cite{nguyen2023hyenadna}, a compact gene language model pretrained at single-nucleotide resolution on the human reference genome~\cite{nihHomoSapiens}. The overall pipeline is illustrated in Figure~\ref{fig:pipeline}. For training, we extracted latent representations from the third layer of HyenaDNA-small-32k using sequences sampled from the human reference genome, each of length 32k nucleotides. To prevent model overfitting to specific genomic contexts, we globally shuffled these activations, ensuring that representations derived from the same input sequence were unlikely to appear together in the same training batch. These processed activations were then used to train SAEs with an expansion factor of 32×, creating feature dictionaries of size 8,192. The learning rate was linearly warmed up over the first 5\% of training steps and then fixed at 1e-6, with an L1 penalty of 0.1 and a batch size of 2,048.

To evaluate whether the resulting sparse features correspond to biologically meaningful genomic elements, we annotated chromosome 14 
with JASPAR transcription factor binding sites (TFBS)~\cite{10.1093/nar/gkad1059}, accessed via the UCSC Table Browser~\cite{10.1093nargkh103}. We then applied quality filtering based on motif frequency and p-value thresholds to retain high-confidence annotations. To facilitate a direct comparison between features and annotations, we converted motif-level annotations into nucleotide-level labels, and used an activation threshold of 0.15 to determine whether an SAE feature was activated or not. This allowed us to systematically assess the alignment between SAE features and known regulatory elements with metrics such as precision, recall, and F1 score, consistent with methodologies introduced by InterPLM~\cite{Simon2024.11.14.623630}.

\section{Results}
\begin{figure*}[ht]
\centering
\includegraphics[width=\textwidth]{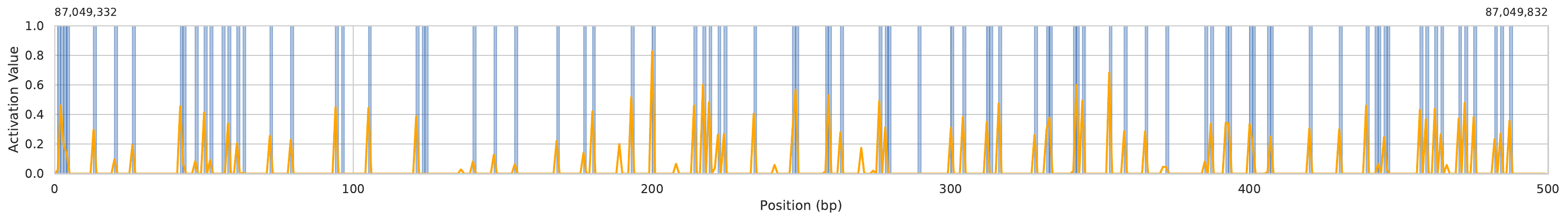}
\caption{Activation pattern of feature f/357 across a 500 bp segment of human chromosome 14. Orange peaks represent activation values, while blue bars indicate cytosine positions starting at position 87,049,332, revealing a strong correlation between the feature and this specific nucleotide.}
\label{fig:act_vis}
\end{figure*}

\begin{figure*}[h!]
\centering
\includegraphics[width=\textwidth]{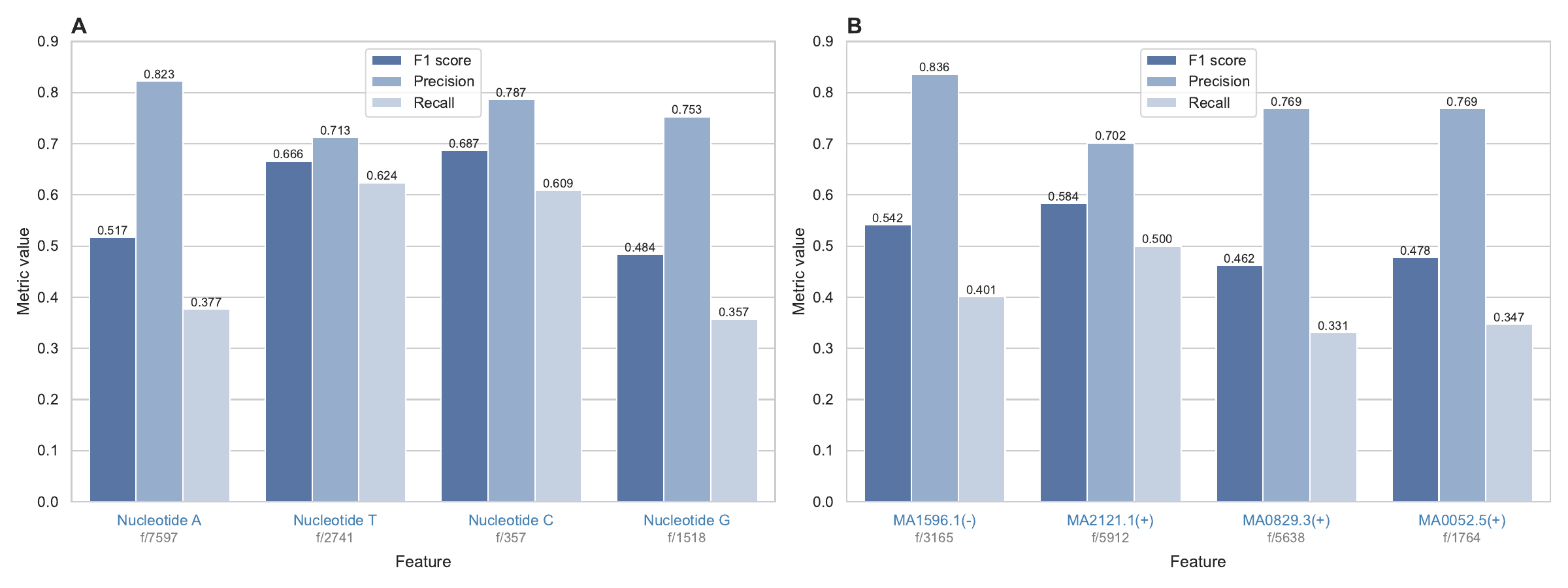}
\caption{Sparse autoencoders reveal interpretable nucleotide and transcription factor binding site (TFBS) features in HyenaDNA-small-32k. (\textbf{A}) Performance metrics for sparse features corresponding to individual nucleotides (A, T, C, G). (\textbf{B}) Metrics for sparse features associated with known TFBSs from JASPAR database~\cite{10.1093/nar/gkad1059}. Strand specificity was indicated by the +/-.}
\label{fig:score}
\end{figure*}

By applying SAEs to embeddings from HyenaDNA-small-32k, we successfully identified sparse features corresponding to individual nucleotides and biologically relevant transcription factor binding sites (TFBS). As shown in Figure~\ref{fig:score}-A, nucleotide-specific features exhibit high precision, which indicates that the learned representations are selective for specific nucleotide identities, although recall varies. This is consistent with prior findings from Evo 2~\cite{Brixi2025.02.18.638918}. Figure~\ref{fig:act_vis} presents the activation pattern of feature f/357 across a 500 bp segment of human chromosome 14. Notably, the activation peaks consistently align with cytosine positions throughout the sequence, demonstrating that this feature has independently learned to recognize this specific nucleotide.

Beyond nucleotide-level features, we identified sparse dimensions aligned with known transcription factor motifs, as illustrated in Figure~\ref{fig:score}-B. Notably, these factors have well-established biological roles: MA1596.1 and MA2121.1 belong to the C2H2 zinc finger factor class, which plays crucial roles in gene regulation. MA0052.5 belongs to the MADS-box class, known for its involvement in muscle development, cell proliferation, and differentiation in animals. The relatively high precision of these features indicates that even compact models capture transcription factor binding specificity effectively. Nevertheless, the observed variability in recall highlights the inherent complexity and redundancy of regulatory elements in genomic sequences.

Overall, we demonstrate that small gene language models encode structured and biologically relevant representations, spanning both nucleotide composition and transcription factor binding patterns.

\section{Conclusion}
Our study demonstrates that sparse autoencoders (SAEs) can extract biologically meaningful representations from small gene language models, revealing structured features at both the nucleotide and regulatory element levels. By applying SAEs to embeddings from HyenaDNA-small-32k, we identified sparse dimensions corresponding to individual nucleotides as well as transcription factor binding motifs, highlighting the ability of compact models to capture essential genomic features. Future research could extend this approach to other genomic contexts, such as non-coding regions or species-specific variations, and explore how SAEs could aid model refinement and interpretability across different architectures. Additionally, SAEs could be applied to other modalities of biological models and data, such as single-cell gene expression and multi-omics datasets, to uncover interpretable representations across diverse biological systems.

\bibliography{biblio}

\end{document}